\documentclass[sigconf]{acmart}
\AtBeginDocument{%
  }
\copyrightyear{2025}
\acmYear{2025}
\setcopyright{cc} 
\setcctype{by}
\acmConference[ICMI '26]{Proceedings of the 28th International Conference on Multimodal Interaction}{October 05--09, 2026}{Napoli, Italy} 
\acmBooktitle{Proceedings of the 27th International Conference on Multimodal Interaction (ICMI '25), October 05--09, 2026, Napoli, Italy} \acmDOI{XXXX}
\acmISBN{XXXX}

\usepackage{enumitem}
\usepackage{booktabs}
\usepackage{threeparttable}
\usepackage{graphicx}
\usepackage{subcaption}
\usepackage{float}
\usepackage{multirow}
\begin{document}

\title  [Educational YouTube Videos and ChatGPT]{How YouTube Frames ChatGPT Use in Education: An Epistemic Network Analysis with Supporting Multimodal Metadata}

\author{Shayla Sharmin}
\email{shayla@udel.edu}
\orcid{0000-0001-5137-1301}
\affiliation{%
  \institution{University of Delaware}
  \streetaddress{South College Avenue}
 \city{Newark}
  \state{Delaware}
  \country{USA}
 }
\author{Mohammad Al-Ratrout}

\email{mratrout@udel.edu}
\orcid{0009-0009-5157-7807}
\affiliation{%
  \institution{University of Delaware}
  \streetaddress{South College Avenue}
 \city{Newark}
 \state{Delaware}
 \country{USA}
 }
\author{Mohammad Fahim Abrar}

\email{fahim@udel.edu}
\orcid{0009-0009-5157-7807}
\affiliation{%
  \institution{University of Delaware}
  \streetaddress{South College Avenue}
 \city{Newark}
 \state{Delaware}
 \country{USA}
 }

\author{Roghayeh Leila Barmaki}
\email{rlb@udel.edu}
\orcid{0000-0002-7570-5270}
\affiliation{%
  \institution{University of Delaware}
 \streetaddress{South College Avenue}
  \city{Newark}
  \state{Delaware}
  \country{USA}
 }

\renewcommand{\shortauthors}{Sharmin et al.}
\begin{abstract}
We examine educational YouTube videos through multimodal metadata, such as transcripts, titles, thumbnails, and viewer comments, to investigate how ChatGPT is framed across creator groups and how those framings relate to audience response and platform reach. Little is known about how large language models are presented to learners in informal, creator-driven public discourse. Following PRISMA, we selected $52$ videos for analysis. We identified three structurally distinct discourse groups: ($G1$) videos that positioned ChatGPT as a conceptual scaffold for thinking, ($G2$) videos oriented toward retrieval practice and skill-building, and ($G3$) videos that framed ChatGPT as a tool for output generation. Epistemic Network Analysis revealed statistically significant group differences with large effect sizes. Multimodal metadata consistently reflected these distinctions across transcript discourse, titles, and thumbnails. Viewers of learning-oriented content described ChatGPT as a thinking partner or tutor, whereas viewers of output-oriented content raised concerns about over-reliance, surface-level learning, and cognitive offloading. $G3$ achieved comparable platform reach to $G2$, yet with substantially weaker learning-oriented framing. This may suggest that output-oriented content competes for visibility despite lower pedagogical depth.
These findings reveal a structural tension in self-directed AI learning: content that prioritizes quick outputs reaches far more learners than content that promotes deep engagement. This gap raises critical questions about whose vision of AI literacy scales and what learners are actually left with.

\end{abstract}



\begin{teaserfigure}
\centering
   \includegraphics[width=\textwidth]{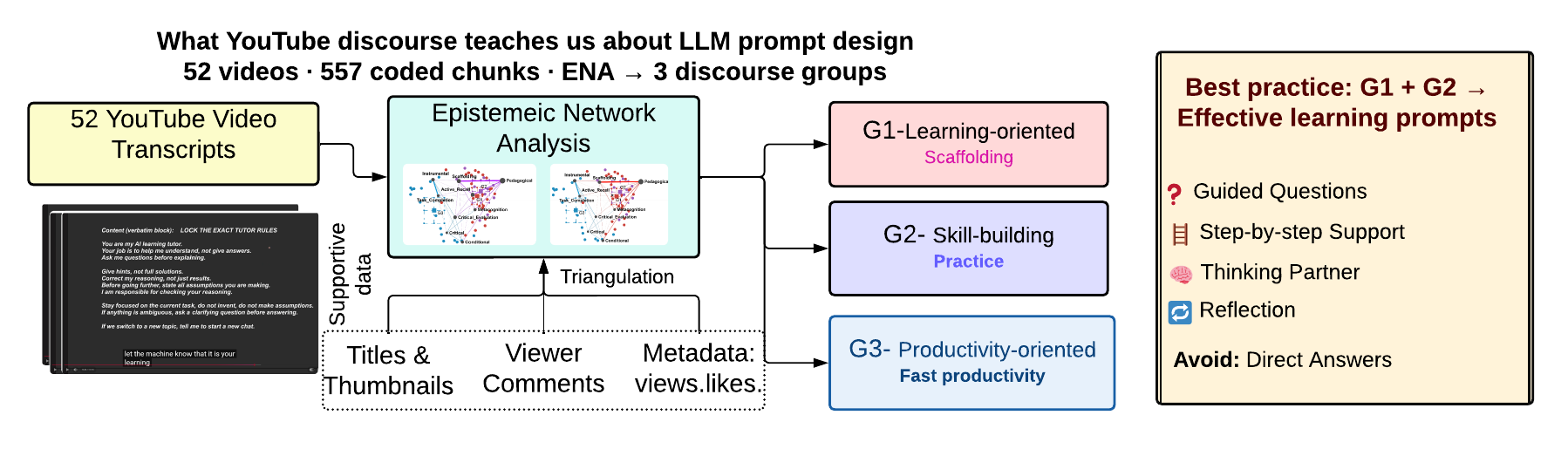}
   \caption{Transcripts from 52 YouTube videos were analyzed using Epistemic Network Analysis, showing three discourse groups: $G1$ (learning-oriented), $G2$ (skill-oriented), and $G3$ (productivity-oriented). For effective learning with LLM, the best practice is to combine the learning and skill-oriented prompts.}
   \label{fig:teaser}
 \end{teaserfigure}


\maketitle
\section{Introduction}\label{sec:Introduction}

The rapid adoption of large language models (LLMs) such as ChatGPT has introduced new questions about how these tools are understood and used in educational contexts. Prior research on AI in education has predominantly examined how tools are introduced within designed instructional environments, such as classrooms and formal curricula, where pedagogical intent is explicit and learner behavior is observable. 
Educational use of LLMs, including ChatGPT, is increasing in formal classrooms, but it is still underexplored how ChatGPT has been framed in informal platforms such as YouTube.
 YouTube videos do more than describe AI use: they package it through discourse, visuals, and platform-facing cues that shape how learners encounter and interpret these tools. This is an informal learning platform, and the creators could be educators or non-educators, but without pedagogical accountability. Also, the audiences are self-selecting, and the platform architecture rewards broad appeal over epistemic depth.
As a result, what learners come to believe about how to use ChatGPT may be shaped more by what the platform promotes than by what actually supports learning.
Despite the scale of this phenomenon, little is known about how ChatGPT is framed in naturalistic, creator-driven public discourse. Specifically, three questions remain underexplored: \textbf{(a)} whether creator framings align with or diverge from learning-oriented approaches, \textbf{(b)} which epistemic orientations achieve the greatest platform reach, and \textbf{(c)} how audiences respond to different framings. To address these gaps, we examine educational YouTube videos as a multimodal site by analyzing creator transcripts, thumbnails, viewer comments, and engagement patterns,  where epistemic norms around ChatGPT are communicated, circulated, and contested. We focus on ChatGPT as it is the most widely adopted and researched LLM to date \cite{Mesko2023ChatGPTRevolution}.
The research questions we asked are as follows:
\begin{enumerate}[label=$RQ_{\arabic*}$]
\item How do different discourse groups frame ChatGPT in educational YouTube videos?
\item How do conditional and critical framings of ChatGPT relate to audience response and platform-level reach in educational YouTube videos?
\end{enumerate}

This study makes three contributions:
\textbf{(1) Discourse structure:} Using epistemic network analysis (ENA), we identify three structurally distinct epistemic orientations in public AI education content. This shows that ChatGPT framing is not a simple learning-to-output binary but reflects coherent discourse constellations. \textbf{(2) Platform dynamics:} We show that platform reach systematically favors output-oriented framing, meaning productivity-focused content reaches more learners than learning-focused content. \textbf{(3) Creator–audience divergence:} Finally, we suggest that audiences make creator messages more explicit by openly discussing and debating them, a pattern less evident in classroom-based research. 
\section{Related Work}\label{sec: Related Work}
\subsection{ChatGPT in Educational Contexts}
ChatGPT has been rapidly adopted in educational settings, and various studies highlight its positive effects on student engagement, critical thinking, and academic achievement due to its dialogic nature \cite{LI2026100571, zhao2023impact}. ChatGPT can be positioned as a tutor because it is capable of providing explanations, personalized feedback, and step-by-step guidance. This \textbf{pedagogical} potential supports students' understanding and skill-building \cite{LI2026100571, DENG2025105224}. 
Research showed that the ability to break down complex ideas has made ChatGPT function as a \textbf{scaffolding} tool \cite{Zhou2025ScaffoldingChatGPT, Guo2025OneYearClassroom} to support learners in ways consistent with Vygotsky's \cite{vygotsky1978} Zone of Proximal Development. 

ChatGPT also supports metacognitive reflection \cite{filiz2025students}, higher-order cognitive performance \cite{Daniel2026ChatGPTIntegration}, and improved exam scores \cite{yusof2025chatgpt} via active-recall/retrieval-practice strategies grounded in the testing-effect literature \cite{roediger2006}.
Additionally, ChatGPT is equally documented as an \textbf{instrumental tool} or productive tool in educational settings \cite{ rasheed2024chatgpt}. This is used for \textbf{task completion}, writing assistance, lesson planning, grading, output generation, and general workload reduction without a learning focus \cite{kosmyna2025your}. 
Researchers have raised concerns about over-reliance and cognitive offloading, where learners outsource cognitive tasks to AI rather than engaging in the processing necessary for durable learning \cite{risko2016}. This has also been seen in \textbf{critical} perspectives on ChatGPT in education. Unrestricted use of ChatGPT could be potentially harmful to conceptual engagement and long-term retention \cite{kosmyna2025your}. Researchers also suggest doing an \textit{evaluation on the responses} by verifying or fact-checking the generated output \cite{holzmann2025chatgpt}.  They also acknowledged that ChatGPT's educational value is \textbf{conditional}, dependent on how it is integrated, monitored, and verified in practice \cite{Wu2026ChatGPTMetaAnalysis}.
Overall, this body of work establishes that ChatGPT can be framed with fundamentally different epistemic roles depending on how it is framed for learners. Most of the research has been done in a classroom \cite{Daniel2026ChatGPTIntegration, yusof2025chatgpt, Guo2025OneYearClassroom, Zhou2025ScaffoldingChatGPT} or with survey teachers \cite{filiz2025students}, how these same framings operate in informal, creator-driven public discourse, and which framings reach learners at scale, remains an open question.

\subsection{YouTube as an Informal Learning Ecology}

YouTube is considered an informal language platform outside formal educational settings without pedagogical accountability specially preferred by adults \cite{quintero2026youtube}.  
Unlike instructional environments with explicit pedagogical design, YouTube is an informal learning ecology shaped by platform logic that prioritizes visibility and engagement over epistemic depth \cite{quintero2024youtube, covington2016}.
A small but growing body of work has begun examining how LLMs such as ChatGPT are discussed in YouTube educational content specifically. 
Folt{\'y}nek and Philip investigated how the YouTube video creators advise students on using ChatGPT \cite{Foltynek2026YouTubeAIBypass}. From 173 videos, they identified two main categories of using ChatGPT as techniques for bypassing AI detection \cite{Foltynek2026YouTubeAIBypass}.
Bal et al. applied opinion mining and sentiment analysis on 66 YouTube videos to analyze content, comments, and transcriptions \cite{bal2024exploring}. They categorized how creators use generative AI for specific skills like speaking and writing.
Anderson and Niu analyzed ``how-to'' YouTube videos to identify common generative AI use cases, such as script editing, translating languages, and correcting speech errors within the videos themselves. 
They showed how LLMs can help in script writing, editing, translating, and correcting speech errors from creating videos to upload online \cite{anderson2025making}.
Overall, these studies establish that YouTube creator discourse around ChatGPT is neither uniform nor pedagogically neutral. However, prior work treats framing as thematic categories rather than examining the structural co-occurrence of epistemic orientations across discourse with multimodal metadata.

\section{Methodology}
\subsection{Search Strategy: Data Collection} \label{sec: Data Collection}
\begin{figure}[hbt!]
    \centering
    \includegraphics[width=\linewidth]{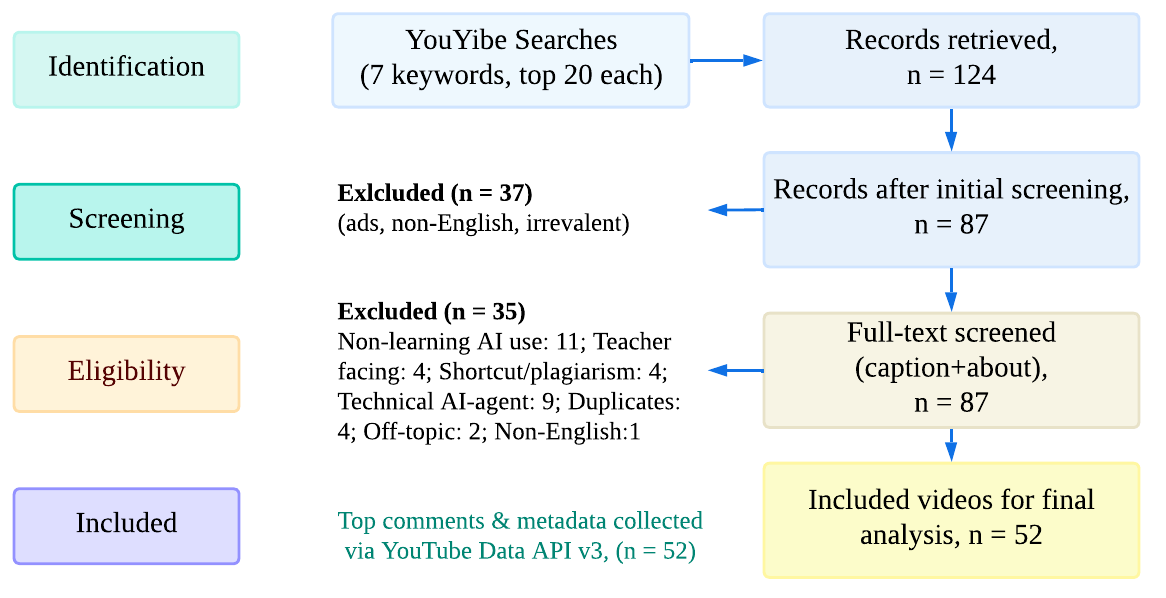}
    \caption{ Overview of the data PRISMA flowchart}
    \label{fig:youtubePrisma}
\end{figure}

We followed ``Preferred Reporting Items for Systematic Reviews and Meta-Analyses (PRISMA)'' ~\cite{page2021prisma} to identify and select videos from YouTube (\autoref{fig:youtubePrisma}). We chose ChatGPT because a high volume of research on ChatGPT has already been conducted, and the results showed a positive effect on students' engagement, critical thinking, and academic achievement in education ~\cite{YOUSSEF2024100316}.
Initially, we finalized seven keywords related to learning and task-oriented framing of ChatGPT in education. Our keywords are as follows:  
``ChatGPT as a tutor'', ``ChatGPT explain to me topic'', ``ChatGPT study partner'', ``ChatGPT as a tutor homework'', ``ChatGPT do my homework'', ''finish assignment with AI'', and ``ChatGPT complete my work''. We searched on February 22, 2026. We planned to collect the top 20 videos from each search.
The initial YouTube searches were conducted in Incognito mode to reduce personalization. 
We chose videos sorted by relevance, length having 3-20 minutes, and uploaded recently (current year).
Two researchers jointly conducted the preliminary screening, and 124 videos were retrieved because some keywords (``ChatGPT do my homework'', ``finish assignment with AI'', and ``ChatGPT complete my work'') returned fewer than 20 results. During this stage, we had 87 videos after removing advertisements, sponsored content, non-English videos, and clearly irrelevant results. Videos were filtered to those published between January 2025 and February 2026 (using ``current year'' filter) to capture recent creator discourse on ChatGPT in educational contexts. We extract video metadata that includes title, channel name, view count, like count, comment count, duration, and publication date using the YouTube Data API v3 via a custom Python script.

In the second screening stage, we reviewed each video's \textit{Caption} and \textit{About} sections. Videos were included if they focused on learners' use of ChatGPT for educational purposes. Videos were excluded if they primarily addressed general AI productivity, business applications, work automation, tool features, or tool migration rather than learning ($n = 11$); teacher-facing rather than learner-facing uses ($n = 4$); shortcut-oriented completion, plagiarism, or AI-detection avoidance ($n = 4$); technical AI-agent coursework unrelated to learner use of LLMs ($n = 9$); duplicate entries ($n = 4$); off-topic or irrelevant content ($n = 2$); or non-English videos ($n = 1$). 
This process resulted in a final dataset of $52$ videos for in-depth analysis with an average length of 7 minutes 47 seconds. 
We also collected the viewers' top comments and metadata for each of these $52$ videos using the YouTube Data API v3 via custom Python scripts. 

\subsection{Transcript Processing and Chunking} \label{sec: Transcript}
After finalizing our videos, we downloaded the transcripts using an automated YouTube transcript extraction tool, NoteGPT (notegpt.io). The exported transcripts were short-time-based units rather than sentence-level utterances. We saved one sentence per row in a CSV file based on punctuation using a customized Python script. 
Then we converted the sentences into meaningful analytical units (chunks) having 3 to 5 sentences. We applied a boundary protocol to create each meaningful chunk. Our structure was to consider a new chunk began when a clear topical shift occurred, often signaled by discourse markers such as ``Now,'' ``Next,'' ``However,'' or ``But.'' Short filler utterances (e.g., ``Let's get started'') were merged with adjacent chunks rather than treated as separate analytical units. Initially, we instructed. Later, we manually checked, reviewed, and corrected all the chunks. The final dataset comprised 557 chunks across 52 videos.

\subsection{Coding Scheme} \label{sec: Coding Scheme}
Prior research has framed ChatGPT as a pedagogical support tool to provide explanations, feedback, personalization, and step-by-step assistance for learners and as an instrument for productivity, output generation, writing support, and task completion. At the same time, scholars have raised critical concerns regarding over-reliance, misinformation, reduced conceptual engagement, and dependency, while others argue that the value of ChatGPT is conditional on how it is integrated, monitored, and verified in practice.
So, we considered two different coding schemes (details in ~\autoref{tab:codes_combined}): \textit{framing} (e.g., pedagogical, instrumental, critical, and conditional) and \textit{strategy} (e.g., scaffolding, active recall, task completion, evaluating response, and metacognition) dimensions in educational YouTube discourse about ChatGPT. These nine binary codes were developed inductively from the data and aligned with the existing framework in educational technology research. 
A subset (10\%) of the data was independently coded by two raters using a predefined binary coding scheme (1 = present, 0 = absent) across nine categories. Inter-rater reliability was evaluated using Cohen’s kappa, which indicates substantial agreement ($\kappa > 0.70$). Discrepancies were resolved through discussion (See Appendix A). Later, the remaining data were coded by one rater and reviewed by the second rater to ensure consistency. An AI-based tool was used to assist with interpretation during coding; however, coding decisions were finalized by the human raters.

\begin{table*}[h]
\centering
\caption{Coding Scheme Definition and Percentage of Coded Chunks by Group. Framing codes are mutually exclusive and sum to approximately 100\% per group (minor deviations due to rounding); Strategy codes are not mutually exclusive, as multiple strategy codes may apply to a single chunk, so columns do not sum to 100\%.}\label{tab:codes_combined}
\footnotesize
\begin{tabular}{lp{8cm}rrr}
\toprule
\textbf{Code} & \textbf{Definition} & \textbf{G1} & \textbf{G2} & \textbf{G3} \\
\midrule

\multicolumn{5}{c}{\textbf{Framing Codes}} \\
\midrule
Pedagogical & ChatGPT used as a tutor to support understanding and skill-building. 
& 64.5 & 74.3 & 2.3 \\

Instrumental & ChatGPT used as a productivity tool to generate outputs without a learning focus. 
& 10.7 & 8.8 & 56.7 \\

Conditional & ChatGPT is beneficial only when used correctly or responsibly. 
& 22.0 & 15.9 & 21.6 \\

Critical & ChatGPT is framed as harmful or damaging to learning or cognition. 
& 3.8 & 0.9 & 19.3 \\

\midrule


\multicolumn{5}{c}{\textbf{Strategy Codes}} \\
\midrule
Scaffolding & ChatGPT breaks down complex ideas step-by-step to support learning. 
& 47.3 & 44.2 & 13.5 \\

Metacognition & Learner monitors and evaluates their own thinking, understanding, and process.
& 26.7 & 23.0 & 17.0 \\

Active\_Recall & ChatGPT is used to test knowledge through quizzes or retrieval practice. 
& 13.7 & 34.5 & 1.2 \\

Task\_Completion & ChatGPT generates completed outputs without learning intent. 
& 10.3 & 8.8 & 34.5 \\
Evaluating\_Response & ChatGPT output is verified or fact-checked by the learner. 
& 12.8 & 16.8 & 20.5 \\
\bottomrule
\end{tabular}
\end{table*}

To ensure coding consistency, each code was accompanied by explicit inclusion and exclusion criteria with annotated examples. Several boundary cases required particular attention. \textit{Instrumental} was not coded if the chunk also
encouraged reflection or verification of output. \textit{Scaffolding} was distinguished from task efficiency; a chunk that divided a report into sections for productivity was not coded as Scaffolding unless understanding was being checked. \textit{Critical} required the warning to be the dominant message of the chunk, not a brief disclaimer. Similarly, \textit{Conditional} required the qualification to be central, not incidental. \textit{Metacognition} was distinguished from task planning; awareness of the learning process itself was required, not merely organizing steps.
To ensure mutual exclusivity of framing codes, each chunk was assigned the most dominant framing orientation. For instance, if a statement contained both conditional and instrumental language, it was coded based on its primary objective: \textit{Conditional} was assigned if the focus was on setting a boundary for responsible use, whereas \textit{Instrumental} was assigned if the statement's primary aim was optimizing output efficiency. A complete coding manual including full definitions, inclusion/exclusion criteria, annotated examples for each of the nine codes, and boundary case decision rules is provided in Appendix A.
\subsection{Group Classification}\label{sec:group-classification}

Videos were categorized into three groups based on their dominant communicative orientation. 
~\autoref{tab:codes_combined} presents the code frequency distribution across groups.
\textbf{Group 1} (G1, \textit{Explanatory/Tutor-like}, $n=25$) comprised videos that framed ChatGPT as a conceptual learning agent. This group was characterized by high Pedagogical (64.5\%) and Scaffolding (47.3\%) code frequencies. 
\textbf{Group 2} (G2, \textit{Practice/Skill-building}, $n=10$) similarly emphasized Pedagogical framing (74.3\%) but was distinguished by elevated Active Recall usage (34.5\%), reflecting a practice-oriented discourse. 
\textbf{Group 3} (G3, \textit{Output/Productivity}, $n=17$) represented a fundamentally distinct orientation, dominated by Instrumental framing (56.7\%) and Task Completion (34.5\%), with Pedagogical framing nearly absent (2.3\%).

\section{Analysis}

\subsection{\textbf{Epistemic Network Analysis Setup}}
In this study, we applied Epistemic Network Analysis (ENA)
 to our data using the ENA
Web Tool (Version 1.8.0) \cite{shaffer2017, shaffer2016, bowman2021, marquart2021}. ENA constructs weighted networks for each unit of analysis, where nodes represent codes and edge weights reflect binary co-occurrence (presence/absence) within a moving stanza window \cite{bowman2021, shaffer2016}. Units of analysis were individual videos ($n = 52$), nested within group (\texttt{group\_code} $>$ \texttt{video\_id}). Each video was treated as a separate conversation, ensuring that co-occurrence connections did not 
accumulate across videos. All nine binary codes served as nodes. A moving stanza window of size $4$ was applied, meaning co-occurrences were calculated across each chunk and the three preceding chunks within the same video \cite{ruis2019, siebert2017}. This window size was selected to capture topically coherent discourse sequences while remaining sensitive to local patterns.

Networks were aggregated using binary summation, in which co-occurrence reflects the presence or absence of each code pair within the stanza window. Networks were normalized before dimensionality reduction to account for differences in the number of coded chunks across videos \cite{bowman2021, behdokht2024Nursing,behdokht2025Mena,behdokht2026Caregiver,shayla2025ENA}. For dimensionality reduction, singular value decomposition (SVD) was applied, producing orthogonal dimensions that maximize the variance explained by each dimension \cite{behdokht2024Nursing,behdokht2025Mena,behdokht2026Caregiver,shayla2025ENA}. The model had co-registration correlations of $0.97$ (Pearson) and $0.97$ (Spearman) for
SVD1, and $0.96$ (Pearson) and $0.96$ (Spearman) for SVD2, indicating strong goodness of fit between the visualization and the original model. Group-level mean networks were compared statistically using Mann-Whitney $U$ tests on projected axis scores \cite{shaffer2016}. Group classification preceded ENA and was based solely on dominant code frequencies in the raw coded data, independent of network-level features. 
ENA was then applied to characterize structural co-occurrence patterns within each group \cite{shaffer2016}.

\subsection{Discourse Patterns Analysis}
\label{sec:content-analysis}

To support the ENA structural findings, we conducted a 
directed qualitative content analysis \cite{mayring2000} of transcript 
chunks, viewer comments. title, and thumbnails as complementary multimodal evidence. Rather than generating new themes inductively for the transcript chunks, this analysis examined how the nine coding categories 
were discursively constructed in creator language, using 
representative excerpts to illustrate the epistemic orientations identified through ENA.

\subsubsection{\textbf{Transcript Analysis}}
All 557 coded chunks across 52 videos were examined in full through iterative review. Recurring discourse patterns within each coding category were identified, and representative excerpts were selected to illustrate how creators framed ChatGPT use through language.

\subsubsection{\textbf{Viewer Comment Analysis}}
A total of 936 comments were collected 
($n_{\text{G1}} = 56$, $n_{\text{G2}} = 306$, $n_{\text{G3}} = 574$).  Given the highly unequal distribution across groups and the concentration of high-engagement $G3$ comments within a single video, comment data were treated as an independent qualitative triangulation layer rather than a basis for cross-group quantitative comparison. 
Each comment was matched to its source video and assigned to a group using the video label mapping created during data collection. 
The comments were then qualitatively reviewed to examine how viewers responded to the discourse patterns reflected in each group’s content. We got comments from 28 videos.
As comments were unavailable for all 52 videos due to disabled comment sections or API limitations, they were used as an independent qualitative layer to triangulate the epistemic orientations, with particular attention to how audiences responded to creator framing across groups.

\subsubsection{\textbf{Video Metadata Analysis}}
For metadata, we collected information such as title, video ID, publication date, total number of views, likes, comments, duration, channel ID, channel title, and the about section. We checked the data manually to understand the relative visibility and engagement across the three video groups. We used a normalized measure for descriptive comparison because raw YouTube counts (views, likes, comments) are highly skewed and increase over time. Video visibility was defined as \textit{views per day}, and engagement was defined as \textit{likes per 1,000 views} and \textit{comments per 1,000 views}.
Median values (Mdn) and interquartile ranges (IQR) are reported because these measures are less affected by extreme values than raw totals.

\subsubsection{\textbf{Video Thumbnail and Title Analysis}}
Thumbnail images were collected and analyzed manually as a video-level visual framing layer. Thumbnail features were coded using a predefined coding scheme that captures visual elements (e.g., presence of person, study context, learning imagery, productivity imagery, AI symbol).
Title text was coded for the presence of learning-oriented, urgency-based, and productivity-oriented language. In both cases, each feature was treated as a binary variable (0 = absent, 1 = present), and group-level percentages were calculated.
One video was unavailable at the time of thumbnail retrieval and was excluded from thumbnail-specific counts while retained in the main dataset.

\section{Results}\label{sec:Results}

\subsection{ENA Results}
Although Table 1 shows that 
G1 and 
G2 have similarly high Pedagogical frequencies (64.5\% and 74.3\%, respectively), frequency distributions alone cannot distinguish these groups' underlying discourse structure. ENA reveals that they differ in which codes co-occur with Pedagogical framing: G2 shows a distinct Pedagogical–Active\_Recall connection ($mean_weight = 0.30$) that is comparatively weak in
G1, providing analytical value beyond what thematic coding frequencies alone can offer.

\begin{figure*}[h]
    \centering
    \includegraphics[width=0.9\textwidth]{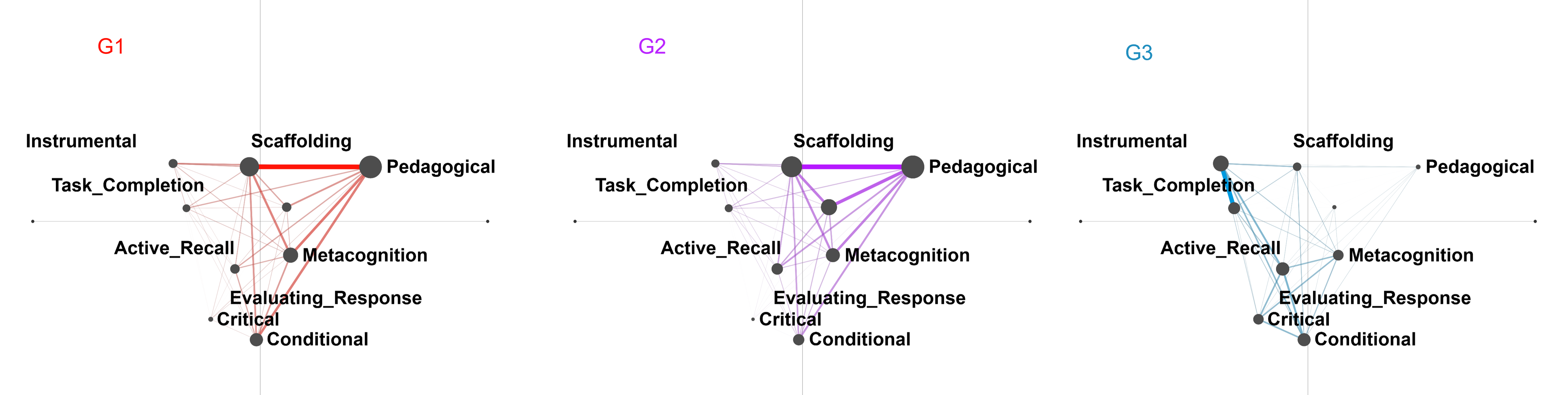}
    \caption{Group-level mean ENA networks for the three discourse groups. Node size indicates code frequency, edge thickness indicates co-occurrence strength, illustrating the distinct epistemic orientations of $G1$, $G2$, and $G3$.}
    \label{fig:g1_g2_g3_individual}
\end{figure*}
\begin{figure*}[h]
    \centering
    \includegraphics[width=0.8\textwidth]{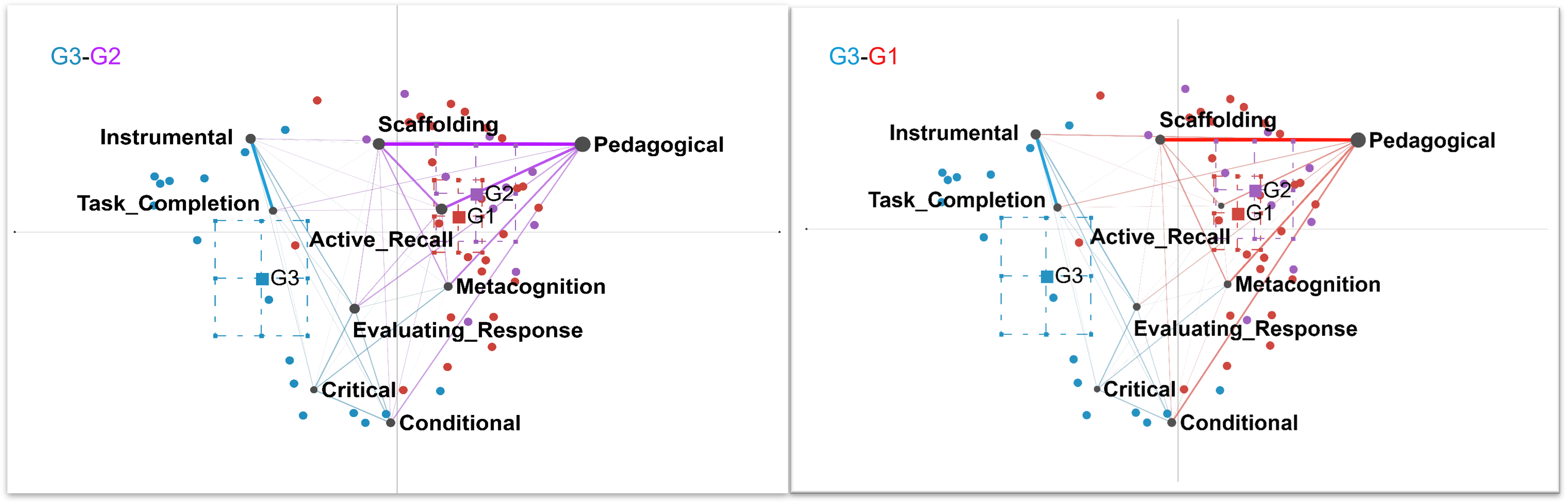}
    \caption{Pairwise ENA centroid and network comparisons across discourse groups. The left panel compares $G3$ and $G2$, and the right panel compares $G3$ and $G1$. Differential edge patterns indicate differences in code co-occurrence structure across groups.}
    \label{fig:ena_comparison}
\end{figure*}

The mean ENA network for $G1$ was anchored by a strong connection between \textit{Scaffolding} and \textit{Pedagogical} ($mean weight = 0.43$), the dominant edge in the network. \textit{Pedagogical} co-occurred consistently with both \textit{Metacognition} ($0.23$) and \textit{Conditional} ($0.22$), and \textit{Scaffolding} co-occurred with 
\textit{Metacognition} ($0.20$) and \textit{Conditional} ($0.16$). \textit{Instrumental} and \textit{Task\_Completion} were near-absent in the $G1$ network (\autoref{fig:g1_g2_g3_individual}(a)).
The mean ENA network for $G2$ shared the strongest connection observed across all groups: \textit{Scaffolding}--\textit{Pedagogical} ($mean weight = 0.42$). However, $G2$ was distinguished by a prominent \textit{Pedagogical}--\textit{Active\_Recall} connection ($0.30$) and a \textit{Scaffolding}--\textit{Active\_Recall} connection ($0.23$), reflecting the retrieval and practice orientation of this group. \textit{Metacognition} ($0.20$) and \textit{Conditional} ($0.18$) remained 
present but comparatively weaker than in $G1$ (\autoref{fig:g1_g2_g3_individual}(b))
The dominant connection for $G3$ was \textit{Instrumental}--\textit{Task\_Completion} ($mean weight = 0.40$). Secondary connections clustered among \textit{Instrumental}, \textit{Critical}, \textit{Conditional}, and \textit{Evaluating\_Response} (weights ranging from $0.14 to 0.17$), reflecting occasional risk acknowledgment. \textit{Pedagogical} and \textit{Active\_Recall} were near-absent nodes in the $G3$ network (\autoref{fig:g1_g2_g3_individual}(c)).
Both $G1$ and $G2$ were significantly different from $G3$ on the SVD1 axis ($p < .001$), with large effect sizes ($r = -0.92$ and $r = -0.95$ respectively). On the SVD2 axis, $G2$ differed marginally from $G3$ ($p = .05$), while $G1$ did not ($p = .07$). 

\subsection{Discourse Patterns from Transcripts}

Discourse-level illustration of 557 transcript chunks across 52 videos identified five discourse patterns that illustrate the coding categories and corroborate the epistemic distinctions revealed by ENA.

\textbf{Discourse 1: ChatGPT as Thinking Partner.}
The most dominant theme across   $G1$ and   $G2$ was how creators framed ChatGPT as an active learning companion rather than an output generator.   $G1$ creators explicitly positioned learners as agents who use ChatGPT to refine their thinking: \textit{``Your goal is to use AI as a wise tutor who guides you with questions, not as a vending machine that just spits out the answer''} (4CT).   $G2$ creators similarly foregrounded practice and engagement: \textit{``I just treated ChatGPT like an English-speaking partner. I talked about my day, how it went''} (2CE). $G3$ creators, by contrast, foregrounded efficiency and output generation: \textit{``stop wasting time asking more and more questions or giving it further feedback and get straight to the point''} (11CT). 

\textbf{Discourse 2: Scaffolding as Central Strategy.}
Scaffolding was explicitly and pedagogically framed in   $G1$ and   $G2$, but was replaced by task efficiency in   $G3$.   $G1$ creators described structured breakdown as a learning process: \textit{``Tutoring is asking better questions, breaking confusion into steps, slowing thinking down''} (2CT).   $G3$ creators used structurally similar language, but oriented toward output rather than understanding: \textit{``This very specific and detailed prompt is going to help generate meaningful questions so it can uncover how you think, how you communicate, and make decisions
the best. We want to customize ChatGPT for your specific needs so you can stop wasting time asking more and more questions or giving it further feedback and get straight to the point''} (11CT). 

\textbf{Discourse 3: Metacognition and Learner Agency.}
Metacognitive prompting was most explicit in   $G1$, with creators
consistently asking learners to assess their prior knowledge before
engaging with ChatGPT: \textit{``Before you start typing anything into
AI, you have to build some context. What do you already know?''} (1CT).
  $G2$ showed practice-oriented metacognition, with creators encouraging
learners to monitor their own progress: \textit{``I started recording
conversations with ChatGPT so that I can listen to myself, notice my
mistakes''} (2CE). In   $G3$, self-reflection was framed around output
quality rather than learning depth, with creators prompting users to
evaluate whether the AI's responses met their productivity needs rather
than deepened their understanding.


\textbf{Discourse 4: Fact-Checking and Verification.}
  $G1$ demonstrated the strongest framing of learners as responsible
evaluators: \textit{``The responsibility is on you. The way you learn
dictates the quality of your learning''} (1CT).   $G2$ showed
practice-oriented awareness, with creators emphasizing verification
alongside engagement: \textit{``Review your chats later and write down
new words or expressions in your notebook --- make sure to go to that
notebook and revise, go through again and again''} (2CE).   
In $G3$, verification appeared but was framed around output quality rather than epistemic accuracy. The creators encouraged checking whether the AI met task requirements, not whether the content was factually correct:
\textit{``Get into the habit of thinking about AI as an assistant rather
than your actual brain --- use it to find resources, as a sounding
board, or to get a quick overview that helps your brain start
thinking''} (6CTH).

\textbf{Discourse 5: Conditional Use and Ethical Boundaries.}
Conditional framing was present across all three groups, but served
qualitatively different functions. In   $G1$ and   $G2$, conditionality
functioned as an epistemic boundary: creators used it to define the
limits of responsible AI use in terms of learning integrity.
\textit{``If AI is easier than thinking, then you must stop the
process''} (1CT) exemplifies this orientation, where the learner's
own cognitive effort is the benchmark. In contrast, while the 'Conditional' code is also present in $G3$, it functions instrumentally rather than pedagogically. Here, creators use conditionality as a prompt engineering heuristic to maximize output efficiency rather than to preserve learning. For instance, the statement, \textit{``Want to know why 90\% of people are getting garbage results from AI? They're missing one critical step''} (3CE), illustrates this 'if-then' logic aimed at productivity optimization. Thus, unlike in $G1$ and $G2$, where conditionality protects the learning process, in $G3$, it serves as a practical tip to improve AI performance.

\subsection{Viewer Comment Pattern}

Thematic analysis of 936 viewer comments across 28 videos identified five themes that further triangulate the ENA findings. Because comment volume was uneven across groups, with many $G3$ comments concentrated in a single high-engagement video, these findings are exploratory and should be interpreted cautiously.
 
\textbf{Discourse 1: AI as Learning Partner vs.\ Shortcut.}
Viewers of   $G1$ and   $G2$ content described ChatGPT as a genuine learning
tool: \textit{``I use it to guide me in learning stuff\ldots{} It's
like having a tutor''} (  $G1$, Likes: 21). In contrast, the most highly liked comments on   $G3$ videos expressed concern about cognitive
offloading: \textit{``It will destroy your brain if you consistently
ask it to do the job for you''} (  $G3$, Likes: 1301) and \textit{``I
noticed within a week that the code problems I overcame did not teach
me anything when assisted by AI''} (  $G3$, Likes: 981).
 
\textbf{Discourse 2: Over-reliance and Cognitive Offloading.}
Over-reliance concern was concentrated in   $G3$ comments, with some of
the highest engagement across the dataset: \textit{``As a teacher for over 30 years --- students have seemingly lost the ability to decide
when something is good enough''} (  $G3$, Likes: 2128). This theme was
absent from   $G1$ and   $G2$ comments, mirroring the Critical framing pattern
in the ENA networks.

\subsection{Video Metadata Framing }
\begin{table}[ht!]
\centering
\caption{Descriptive visibility and engagement metrics by group. IQR=interquartile ranges,
Mdn = median; values in parentheses indicate number of videos (n).}
\label{tab:metadata}

\footnotesize

\begin{tabular}{llll}
\toprule
Metric &   $G1$ (n=25) &   $G2$ (n=10) &   $G3$ (n=17) \\
\midrule

Views/day (Mdn) & 0.27 & 20.87 & 14.34 \\
Views/day (IQR) & [0.06, 0.89] & [0.69, 548.87] & [0.25, 460.38] \\

Likes/1k (Mdn) & 46.51 & 48.63 & 30.28 \\
Likes/1k (IQR) & [10.87, 63.51] & [1.35, 99.40] & [14.46, 49.22] \\

Comments/1k (Mdn) & 0.00 & 0.75 & 1.07 \\
Comments/1k (IQR) & [0.00, 11.33] & [0.00, 2.63] & [0.00, 5.32] \\
\bottomrule
\end{tabular}
\end{table}

~\autoref{tab:metadata} presents descriptive visibility and engagement metrics by group.   $G2$ showed the highest typical visibility, with a median of $20.87$ views per day, followed by   $G3$ $(14.34)$ and   $G1$ $(0.27)$. Median likes per 1,000 views were similar in   $G1$ $(46.51)$ and   $G2$ $(48.63)$, but lower in   $G3$ $(30.28)$. Median comments per 1,000 views were highest in   $G3$ $(1.07)$, followed by   $G2$ $(0.75)$, while   $G1$ showed minimal comment activity. The IQR values show that   $G2$ and   $G3$ had much wider variation in visibility than   $G1$. This means that some videos in these groups performed much better than others, while   $G1$ videos were more consistently low in visibility.

\subsection{Thumbnail and Title Framing}

Based on 51 of the 52 videos (one video's thumbnail was unavailable at the time of retrieval), title analysis shows differences in linguistic framing across groups.   $G1$ (68.0\%) and   $G2$ (70.0\%) most frequently used learning-oriented languages, whereas   $G3$ showed substantially lower use (29.4\%). In contrast, urgency language was highest in   $G3$ (41.2\%), compared to   $G1$ (20.0\%) and   $G2$ (10.0\%). Productivity-oriented language was also most prominent in   $G3$ (76.5\%), followed by   $G2$ (40.0\%) and   $G1$ (12.0\%).
Thumbnail analysis showed parallel differences in visual framing.   $G1$ exhibited the highest rates of learning imagery (64.0\%) and study context (48.0\%), with   $G2$ showing a similar but less pronounced pattern (40.0\% and 30.0\%, respectively). In contrast,   $G3$ showed substantially lower learning imagery (17.6\%) and study context (11.8\%), but the highest rate of productivity imagery (29.4\%). 
Additional visual cues further differentiated groups. Speaker presence was most frequent in   $G2$ (90.0\%), compared to   $G1$ (56.0\%) and   $G3$ (47.8\%). AI symbolism appeared more often in   $G1$ (52.0\%) and   $G3$ (35.3\%) than in   $G2$ (20.0\%).

\begin{figure}[ht]
    \centering
    
    \label{fig:thumbnail_framing}
    \includegraphics[width=\linewidth]{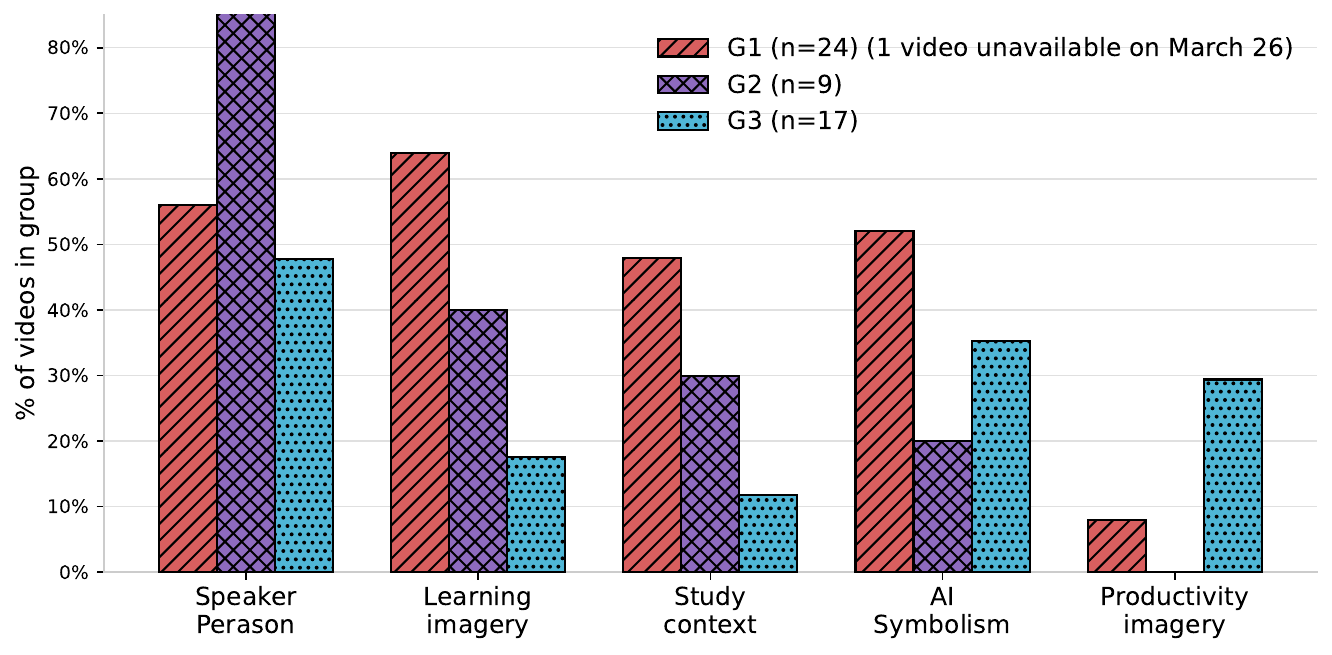}
    \caption{Selected thumbnail framing cues across the three discourse groups ($n=51$; one thumbnail unavailable). Learning-oriented cues were more common in $G1$ and $G2$, whereas $G3$ emphasized productivity imagery and greater visual intensity. Percentages are within-group.}
\end{figure}

\section{Discussion}

\subsection{$RQ_1$. Epistemic Framings of ChatGPT Across Discourse Groups}
Our results show that educational YouTube videos do not frame ChatGPT uniformly. All three groups positioned ChatGPT as a learning tool but with three distinct frames: a scaffold for thinking ($G1$), a support for practice ($G2$), and a direct engine for producing answers and outputs ($G3$). $G1$ is learning-oriented and $G2$ is skill/practice-oriented; both differ from $G3$'s output focus, though $G1$ and $G2$ also hold subtly different epistemic positions. This distinction is not merely descriptive but structurally meaningful: a denser Scaffolding–Pedagogical cluster in $G1$ and $G2$ suggests creators framed ChatGPT not merely as a tutor, but as a guided learning partner requiring reflection and responsible use, while $G2$ additionally promotes active skill-building. This extends prior work on AI tool framing in education \citep{selwyn2019, roll2016} by showing that the learning-to-output spectrum is not a simple binary but is structured through distinct constellations of co-occurring practices. $G3$, by contrast, presents ChatGPT as a productivity tool in which learners act as recipients of output rather than agents of learning.

A key difference among $G1$, $G2$, and $G3$ lies in the positioning of learner agency. The Scaffolding–Metacognition–Conditional cluster in $G1$ suggests guided breakdown paired with learner reflection, consistent with Vygotsky's \citep{vygotsky1978} Zone of Proximal Development, in which scaffolding gradually transfers cognitive responsibility to the learner. $G2$ shares this orientation but redirects it toward retrieval practice via a distinct Pedagogical–Scaffolding–Active\_Recall connection, aligning with the testing effect \citep{roediger2006} and suggesting that $G2$ creators draw—explicitly or not—on evidence-based strategies; however, weaker Metacognition and Conditional connections than $G1$ suggest this practice orientation is less explicitly reflective. $G3$'s near-absence of Scaffolding, Metacognition, and Pedagogical connections is the more striking structural finding: ChatGPT is positioned not as a scaffold that gradually withdraws, but as a substitute for cognitive effort, paralleling concerns in the cognitive offloading \citep{risko2016} and AI over-reliance \citep{kosmyna2025your} literatures. Whether this framing actually affects learner processing, however, was not examined in this study.

The multimodal findings support the ENA results: groups that presented ChatGPT as a learning support tool in transcripts also framed it in titles and thumbnails around studying and guided learning, whereas the more productivity-oriented group used titles and thumbnails emphasizing speed and efficiency. These differences are thus consistently reflected across verbal, textual, and visual presentation, indicating that the three orientations form coherent epistemic constellations rather than a simple learning-to-output binary.

\subsection{$RQ_2$. Epistemic Framing, Audience Response, and Platform Reach}

All three groups used conditional language when discussing ChatGPT, but its function differed across contexts. In $G1$ and $G2$, conditionality functioned as an epistemic boundary — for example, creators emphasized that if AI makes a task easier than thinking, the learner should stop, reinforcing that ChatGPT should support rather than replace cognitive effort. In $G3$, by contrast, conditional language was oriented toward productivity optimization, functioning as a practical tip for improving output. Similarly, creators in $G1$ and $G2$ framed critical concerns as a learner obligation — encouraging users to guide the AI, avoid shortcuts, and fact-check responses — whereas $G3$ creators' acknowledgments of AI's limits appeared as isolated statements rather than integrated pedagogical guidance, which may limit their influence on learner behavior.

Viewer comments provide independent qualitative support for this interpretation, though because the majority of high-engagement $G3$ comments originated from a single video, these patterns should be interpreted cautiously as illustrative of possible audience reception rather than representative of group-level response. Notably, $G3$, despite having the least learning-oriented framing, achieved platform reach comparable to $G2$ with markedly higher comment engagement, suggesting that productivity-oriented content is associated with both attention and audience unease. In the most highly engaged $G3$ videos, audiences echoed concerns that creators only briefly acknowledged — a tension between creator framing and audience reception, though not representative of the broader viewing audience. This concentration of high-engagement critical comments suggests that productivity-oriented framing may simultaneously attract attention and generate unease, reinforcing that the epistemic orientations identified in ENA are meaningfully perceived by audiences, not merely present in discourse. This creator--audience divergence would be invisible in controlled classroom settings, where learner responses are shaped by instructional design rather than emerging organically from public engagement.

This pattern is also visible in video metadata: the most learning-aligned group ($G1$) showed relatively lower visibility than $G2$ or $G3$, and greater exposure to $G3$-style framing may be associated with a normalization of cognitive offloading and a shift toward output generation over deep engagement — a risk that viewer comments suggest audiences are often more critically aware of than creators themselves. $G1$'s minimal comment activity is consistent with its lack of strong instrumental framing, while $G2$'s stronger like-based engagement aligns with its practice-oriented framing. This distributional asymmetry highlights a structural tension in the YouTube ecosystem: the most learning-aligned content achieved the lowest visibility, while skill- and productivity-oriented content reached substantially more learners. These findings suggest that naturalistic, creator-driven public discourse operates by a logic that appears associated with platform architecture favoring broad appeal over epistemic depth, though the present design cannot establish this as a causal mechanism — pointing to a dimension of AI educational influence that classroom-based research is structurally unable to capture.

\subsection{Limitations and Future Work}
\label{sec:limitations}
Several limitations should be noted. First, we analyzed creator discourse and viewer comments rather than measuring learning outcomes directly. Second, comments were unevenly distributed across groups, with most highly liked G3 comments coming from a single video, limiting comparisons. Third, the dataset was limited to ChatGPT videos collected within a specific search strategy and time period, so it may not generalize to all LLM-related educational content. Fourth, although the coding scheme showed acceptable inter-rater reliability, it reflects the researchers' theoretical perspective, and other frameworks may produce different results. Finally, group classification and ENA used the same nine codes, so ENA results should be interpreted as describing internal co-occurrence patterns within predefined groups rather than independently confirming group separation. Future work should include other LLMs, add audio features (e.g., tone and speaking speed), use longitudinal and experimental designs to study learning outcomes and AI use over time, and apply clustering on ENA features to independently validate group structure.

\section{Conclusion} \label{sec:Conclusion}
This study examined how educational YouTube videos frame ChatGPT and how these framings relate to audience response and platform reach. We identified three discourse patterns: ChatGPT as a conceptual scaffold, as retrieval practice, and as a productivity tool. Multimodal metadata supported these distinctions. Learning-oriented framings aligned with active cognitive engagement, while productivity framing positioned ChatGPT as a substitute for effort. However, the productivity group achieved reach similar to skill-oriented content, with higher comment engagement and greater view variability, suggesting that platform dynamics—not only creator intent—may shape exposure. As LLMs become part of self-directed learning, their public framing may be as important as their instructional use.

\textbf{Safe and Responsible Innovation Statement:}
 This study analyzed publicly available YouTube videos and viewer comments. No personal data was collected, and no individuals were directly involved in the research. All comment data were treated anonymously, and no attempt was made to identify individual users. We promote transparency in how AI tools are framed in public educational media, and we hope this work supports more responsible and learning-centered use of AI in education.
 \section*{Acknowledgment}
This work was supported by the National Science Foundation (NSF) under grants \#2222661-2222663, \#2321274, and \#2426003. 
 The authors also acknowledge funding from the Institute for Engineering-Driven Health at the University of Delaware and the National Science Foundation Accelerating Research Translation program under award number \#2331440. Any opinions, findings, and conclusions expressed in this material are those of the authors and do not necessarily reflect the views of the funding agency.
\balance

  \bibliographystyle{ACM-Reference-Format}
  \bibliography{bibfile}

\appendix

\end{document}